\documentclass[aip,pof,reprint]{revtex4-1}

\usepackage{amsmath}
\usepackage{amssymb}
\usepackage{graphicx}
\usepackage{dcolumn}
\usepackage{bm}

\begin{document}

%\preprint{}

\title{Filamentary vortex in a swirling turbulent flow under strong stretching} %Title of paper



\author{R. Labb\'e}
\email[]{raul.labbe@gmail.com}
%\thanks{} % Use this in place of foot notes on the title page.
\affiliation{Laboratorio de Turbulencia, Departamento de F\'isica,
Facultad de Ciencia, Universidad de Santiago de Chile, USACH. Casilla
307, Correo 2, Santiago, Chile}

\author{G. Bustamante}
\affiliation{Laboratorio de Turbulencia, Departamento de F\'isica,
Facultad de Ciencia, Universidad de Santiago de Chile, USACH. Casilla
307, Correo 2, Santiago, Chile}

\date{22 May 2014}

\begin{abstract}
Energy dissipation and pressure fluctuations are features inherent to turbulent fluid motion, and the internal origin of spatial and temporal temperature fluctuations. By measuring local velocity, pressure or temperature, it is possible to detect the presence of coherent structures ---like concentrated vortices--- and quantify their intensity. Here we report measurements of speed and temperature in a filamentary vortex lying in a strongly stretched turbulent airflow. The largest core temperature drop measured in our experiment was $\Delta T = 9$~K, while the fastest measured air speed was $v_{max} = 200$~m/s, which corresponds to a Mach number $M_{max} = 0.59$. At $M \approx 0.1$ temperature drops are on the order of $0.5$~K. These results are consistent with the model for compressible vortices by Aboelkassem and Vatistas [J. Fluids Eng. {\bf 129}, 1073 (2007)].
\end{abstract}

\pacs{47.27.Jv 47.40.-x 47.32.C-}

\maketitle

\section{\label{Intro}Introduction}

Vortices can be considered among the most amazing and sometimes frightening natural phenomena. They are ubiquitous: everywhere fluid motion takes place, vortices are likely to be found ---the conditions necessary for their formation are not quite stringent. Their length can go up to scales comparable to the size of the whole volume of fluid where they lie, while their width can become comparable to the small scales where viscous forces start to be preponderant in the dynamics of the fluid motion. At a planetary scale, Jupiter's Big Red Spot is the biggest and oldest known vortex in the atmosphere of a planet.\cite{Ingersoll} The largest spatial scale of this anticyclonic vortex is $L \approx 17,000$~km, with measured tangential velocities up to $v_t = 190$~m/s.\cite{Simon} In the Earth atmosphere, the most intense tropical cyclone observed in the Atlantic basin was Hurricane Wilma in October 2005, whose minimum sea level pressure estimate was $p \approx 882$~hPa. This pressure, the lowest ever seen in a hurricane, correlated with its maximum wind velocity: $v_w = 82$~m/s, measured at the time in which the eye contracted to a diameter $D \approx 3.7$~km.\cite{Pasch} At a lower scale we find tornadoes. They have much smaller diameters: between few meters and up to 4~km at ground level. Nevertheless, their winds can be much stronger, going from $v_\phi \approx 17$~m/s for the weakest up to $v_\phi = 135 \pm 10$~m/s for one of the most intense tornadoes ever recorded: the Oklahoma City F5 tornado which was part of the outbreak of 3 May 1999.\cite{DOW}

For obvious reasons, no much data is available on the pressure profile within tornadoes. A remarkable measurement is a recording evidencing a pressure drop $\Delta p = 100$~hPa in the core of the Manchester F-4 tornado of 24 June 2003.\cite{Samaras} In addition to their intrinsic scientific interest, the harmful potential that these large scale vortices have for human life and infrastructure of populated areas constitute an important motivation to make them a subject of active research.

From a practical point of view, vortical flows are also of relevance in a variety of technological and industrial applications in which fluids play a preponderant role, as well as in basic research in fluid dynamics and turbulence. Consequently, they motivate sustained theoretical and experimental research efforts.\cite{Green} An account of theoretical and experimental works made to model 2D and 3D vortices in a variety of cases, starting from $1858$ with the works by Rankine is given by Aboelkassem and Vatistas.\cite{AboVat2008} For the experimental research of vortices it is advantageous to produce them in the controlled environment of a laboratory. In fact, laboratory vortices also have a long history, and a good account of theory and experiments can be found in the book by Alekseenko {\it et al.}.\cite{Alek2007} A development aimed specifically to laboratory quantification of the effects of tornado like vortices on civil terrestrial structures was reported by Haan {\it et al.}.\cite{HaanSWG2008}
\begin{figure*}[t]
\centering \vspace{0.2cm} \hspace{-0.2 cm}
\includegraphics[width=.95\textwidth]{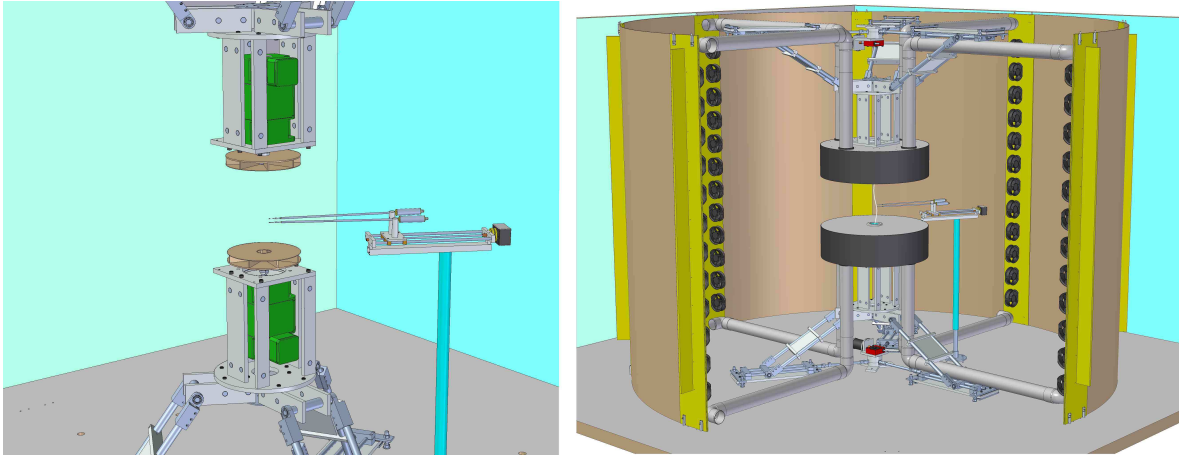}
\caption{{(Color online) (Left) Experimental setup to produce vortices at $M \sim 0.1$
The centrifugal fans rotate in the same direction. The airspeed and temperature probes can be located at varying distances from the flow axis. (Right) Drawing of the room and the device used to produce a filamentary vortex at $M \approx 0.5$ (see text), with the ceiling and some of the walls removed. The vortex core is represented by the white filament at the center, in close proximity to temperature and hot-wire probes. Heath exchangers (not shown) extract the heat produced by the turbines that pump the flow, keeping the air temperature below $33 ^\circ$C.}} \label{one}
\end{figure*}

Here we report experimental results in turbulent vortices produced in air for Mach numbers $M \sim 0.1$ and $M \sim 0.5$. At low Mach numbers it is possible to measure air speed and temperature profiles in a vortex with some degree of detail. For $M \sim 0.5$ measurements are much more difficult, because the vortex becomes a thin filament of diameter $d \sim 1$~mm which is (locally) destabilized by the interaction with the probes. Thus, in the latter case we give air speed and temperature drop measurements near and within the vortex core rather as global changes, without attempting detailed descriptions of velocity or temperature profiles. It is worth to stress here that, in both cases, what we have are {\it free} vortices that share some of the characteristics of the Burgers' vortex,\cite{Burgers48} and are in addition remarkably robust. They do not lose their global structure when locally invasive interactions with the probes take place, although ---as noted before--- a local flow disturbance does occur, particularly in the case of the filament.

In the following section we give a description of the two experimental setups used to produce vortices at low and intermediate subsonic Mach numbers. In Sections \ref{Results:1} and \ref{Results:2} we give the results of measurements of airspeed and temperatute at $M \sim 0.1$ and $M \sim 0.5$, respectively. Finally, in Section \ref{Conc} we discuss our findings in relation to the theory and end with some concluding remarks.

\section{\label{Setup}Experimental Setup}

We used two experimental setups to perform our measurements. As shown in Fig.~\ref{one} (left), the first one consists of two covered centrifugal fans, each facing the other, with central holes to draw the rotating air expelled from their periphery. A detailed description of this device and the measurement system was given in a previous work.\cite{LaBauBus07} In the present case the fans rotate at $f_{\rm rot} = 30$~rps, which gives a Mach number $M \approx 0.1$. The vortex length and diameter are $L = 34$~cm and $d \approx 2$~cm, respectively. The second setup, depicted in Fig.~\ref{one} (right), consists of a cylindrical room with diameter $4.5$~m and height $3$~m. At the center, two ``stretchers'' consisting of vertical cylindrical containers of diameter $D = 92$~cm and height $H = 27$~cm, placed $34$~cm apart, draw the air through central holes of diameter $d = 8$~cm. They enclose two sets of eight two-stage centrifugal turbines, each one driven by an electric motor, that pump the air at an adjustable flow rate. Eight ducts on the opposite sides of the containers canalize the airflow to the outlets located close to the wall of the cylindrical room, where four sets of $12$ fans each, arranged in columns, force the global flow circulation. This setup produces a recirculating flow in which global circulation is generated near the wall at a distance $R \approx 2.25$~m from the room center, where the flow is vertically stretched. The maximum tangential speed of the air near the wall is $v_{\phi} \approx 4$~m/s when the fans rotate at about $80$\% of their maximum capacity. At maximum capacity, the total power consumption of the whole system can reach $P_{\rm max} \approx 16$~kW, but for this experiment the electrical power consumption was $P \approx 8$~kW. A fraction of the total electric power is wasted as heath by the electric motors and transferred to the circulating air, which is used in part as cooling fluid, while the remaining fraction is converted in the mechanical power delivered by the turbines, where another fraction contributes to the air heating through the production of turbulence. The remaining mechanical power is used in producing the mass flow. Eight heat exchangers installed at the outlets of the channelizing ducts remove the heat from the air, keeping the room temperature below $33^\circ$C. Under the specified conditions the airflow through the system is $\Phi \approx 0.8$~m$^3$/s, giving a mass flow $Q \approx 1$~kg/s. The mean airspeed at each stretcher intake is $V_{\rm int} \approx 74$~m/s. By properly adjusting the global flow circulation we obtain a vortex whose Mach number is $M \sim 0.5$. Its exposed length is the same that the separation between the flow stretchers, that is, $l = 34$~cm. Our estimation of the vortex core diameter is $d \sim 1$~mm, which justify its qualification as a filamentary vortex.

It is worth to note that only a small fraction of the angular momentum injected near the room wall is transferred to the vortex. In fact, the air at intermediate distances from the center has a very small speed. We believe that the turbulence generated by the fans enhances the transfer of angular momentum to the center of the room. Thus, keeping all the parameters unchanged, attempting to laminarize the flow produced by the fans could indeed reduce the vortex circulation. The price we pay for a higher circulation at the center of the room is an increased level of turbulence around the filamentary vortex.

The measurements of airspeed were made using a hot wire (HW) anemometer module model $90$C$10$ from Dantec Dynamics, mounted in a Streamline frame. Calibrations were made using a $90$H$02$ Flow Unit. The HW probe, model $55$P$11$, has two conical prongs of length $5$~mm and a separation of $1.4$~mm. Between them a $5 \mu$m diameter platinum plated tungsten wire is connected. Its operating temperature is $300 ^\circ$C. The temperature probe is made of a thermistor (Thermometrics BB05JA102J) with a bead of ellipsoidal shape, having a length of $100 \mu$m and leads with a diameter of $17 \mu$m. These leads were reinforced adding a film of isolating varnish between them, after several occurrences of leads bending by the filament flow. The electronics for temperature measurements was designed and made in our laboratory, and the calibration performed using a commercial K thermocouple based dual digital thermometer with a resolution of $0.1$~K. Several models of digitizing cards from National Instruments, along with the LabView programming software were used to control the experiments and perform the data acquisition.

\section{\label{Results:1}Results for $M \sim 0.1$}

\begin{figure}[t]
\centering \vspace{0.2cm} \hspace{-0.2 cm}
\includegraphics[width=.48\textwidth]{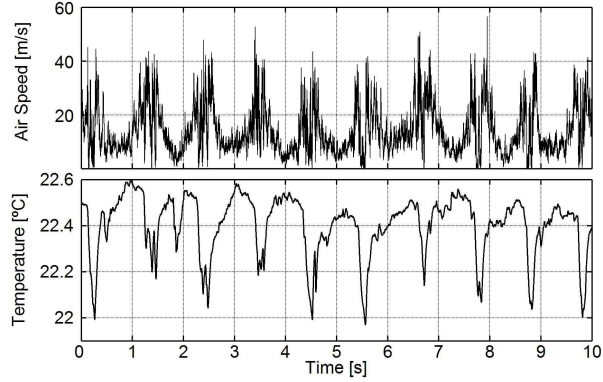}
\caption{Airspeed (top) and temperature (bottom) of a vortex produced between two centrifugal fans. Low temperature
events correlate with the passage of the temperature probe through the vortex core (see text).} \label{two}
\end{figure}

\begin{figure}[t]
\centering \vspace{0.2cm} \hspace{-0.2 cm}
\includegraphics[width=.48\textwidth]{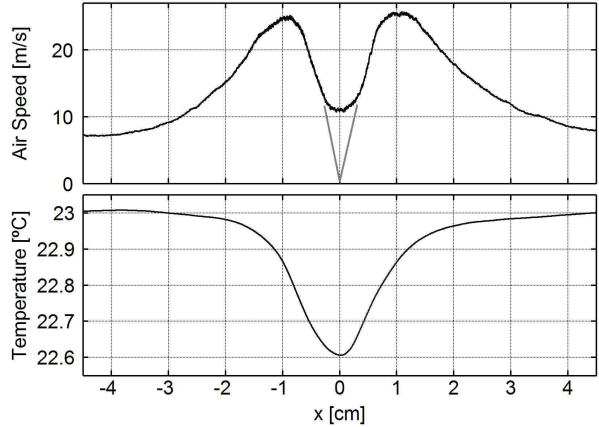}
\caption{Coherent average of airspeed (top) and temperature (bottom) profiles. The horizontal scale,
given in cm, shows that the vortex radius is about $1$~cm. The airspeed profile fails giving zero
speed at the vortex axis, due to the signal blurring produced by the turbulence. The gray lines
give an idea of the true airspeed profile. The slight asymmetry can be ascribed to the change
in the probe orientation with respect to the local velocity direction during the passage of the
vortex.} \label{three}
\end{figure}

Let us see in first place the results for $M \approx 0.1$. Previous works using centrifugal fans show that the vortex produced between them performs a slow, almost periodic precessing motion with a period $T \approx 1$~s for experimental setups having sizes similar to the one used in this work.\cite{LaPiFau1996,LaPi1998,ChiPiLa1996,LaBauBus07} This implies that probes located near to the rotation axis of the fans will produce a fluctuating, almost periodic signal. This is shown in Fig.~\ref{two}, where signals corresponding to airspeed and temperature are displayed. The sampling rate of data acquisition was $15$~kS/s. In this case, both probes were placed at a distance of $1.5$~cm from the rotation axis, which is the approximate precession radius. The vertical separation between probes was $1$~cm. Using the coherent average method described in a previous work,\cite{LaPiFau1996} we can obtain the average airspeed and temperature profiles shown in Fig.~\ref{three}. We can see in Fig.~\ref{three} (top) that the vortex mean radius is $r \approx 1$~cm. The average temperature drop displayed in Fig.~\ref{three} (bottom) is related to air compressibility. It is generally accepted that for $M < 0.3$ compressibility effects can be neglected. In this case we observe in Fig.~\ref{two} (bottom) temperature drops $\Delta T \approx 0.5$~K in the vortex core, while the average temperature drop is, from Fig.~3 (bottom), $\langle \Delta T\rangle = 0.4$~K.
In fact, these values can be considered small, because they represent an air temperature variation of only $0.15$\% with respect to the ambient temperature measured in K. A new analytical solution for self-similar compressible vortices \cite{AboVat2008} gives a temperature drop $\Delta T = 0.43$~K, which is in a remarkable good agreement with our measurements. This theoretical temperature drop is obtained observing that the airspeed peaks in Fig.~\ref{two} (top) have a value around $v_\phi^{\rm max} = 40$~m/s, which corresponds to a Mach number $M = 0.12$. The maximum average value in Fig.~\ref{three} is much smaller, due to the shaking of airspeed peaks by the turbulence within and around the vortex. The final effect is that extreme values, which have a rather low probability of occurrence, are averaged with intermediate speed values which have much higher probability. This effect can be clearly seen in Fig.~\ref{three} (top), where the speed at the vortex axis, which should be nearly zero, is raised by more than $10$~m/s. Of course, the same effect is seen at $x = \pm 1$~m/s, where the maximum airspeed is much smaller than the peaks in Fig.~\ref{two} (top).

\section{\label{Results:2}Results for $M \sim 0.5$}

\begin{figure}[t]
\centering \vspace{0.2cm} \hspace{-0.2 cm}
\includegraphics[width=.48\textwidth]{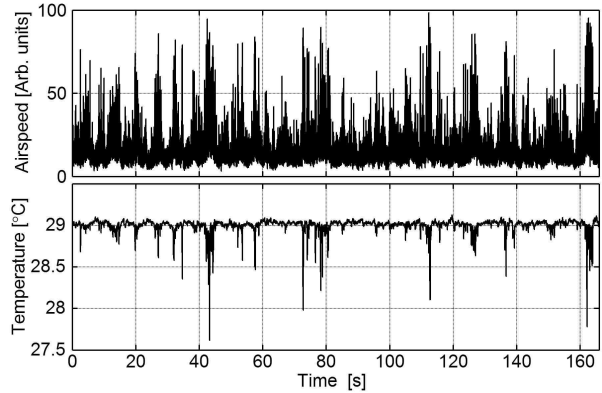}
\caption{Airspeed (top) and temperature (bottom) near the axis of the setup shown in Fig. \ref{one}.
The vortex sweeps the probes at uneven time intervals, as can be seen on both plots. Note that peaks
measured by the airspeed probe have not always corresponding temperature drops (see text).} \label{four}
\end{figure}

Now let us see experimental results obtained at $M \sim 0.5$. Airspeed and temperature probes were successively located at various distances from the axis of the stretching flow. The vertical distance between probes was $\delta_p = 1.5$~mm. Fig.~\ref{four} displays an example of airspeed signal and temperature, as recorded at a location $r_p = 4$~mm from the flow axis, and acquired using a sampling frequency of $12$~kS/s. Although in this plot the speed signal is not calibrated, it is useful to evidence that the nature of signals is very different from those displayed in Fig.~\ref{two}. Here the vorticity filament performs a wandering rather than precessing motion around the flow axis. Thus, it is difficult to evaluate the size of the core from the measurements. As can be seen in Fig.~\ref{four}, only part of the events of maximum local speed coincide with events of low temperature, making an additional difference with the low Mach number case. This is probably due to the small diameter of the filamentary vortex. The size of the probes, although small, are comparable to the size of the vortex core. We can imagine that when the airspeed probe breaks the flow around the filament core, locally destroying the vortex structure, the internal pressure suddenly rises leading to the loss of previously existing temperature, pressure and velocity distributions. Thus, we proceeded performing independent airspeed and temperature measurements to avoid, or at least reduce as much as possible this problem.
\begin{figure}[t]
\centering \vspace{0.2cm} \hspace{-0.2 cm}
\includegraphics[width=.48\textwidth]{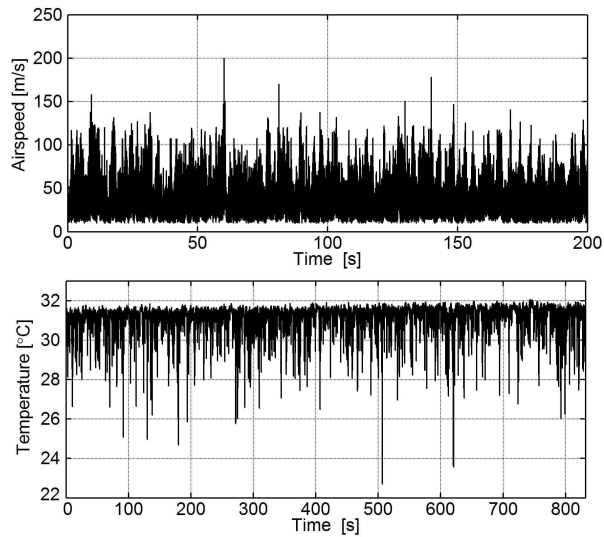}
\caption{Airspeed (top) and temperature (bottom) taken independently at $4$~mm from the flow axis.
Note that the recording time is longer in the lower plot.} \label{five}
\end{figure}

Fig.~\ref{five} shows two independent measurements of airspeed (top) and temperature (bottom). Note the different extents of time scales. In the record of Fig.~\ref{five} (top) we see the occurrence of the maximum speed event in this experiment. Its value was $v^\star = 200$~m/s, corresponding to a Mach number $M^\star = 0.59$. Fig.~\ref{five} (bottom) displays an unrelated and longer record of temperatures. In this case, the largest drop in temperature that we obtained, under the same conditions that produced $v^\star$, was $\Delta T^\ast = (9 \pm 0.2)$~K.

Although we made no attempts to measure the diameter of the filamentary vortex, we used our data to obtain an estimate using a vortex model. Specifically, we used the Burgers vortex,\cite{Burgers48} for which the velocity field in cylindrical coordinates $(r, \phi, z)$ is given by
\begin{equation}
{\mathbf v} = \bigl(-Ar,\frac{\Gamma}{2 \pi r}\bigl(1 - e^{-Ar^2/2\nu}\bigr),2Az\bigr), \label{eq_Burguers}
\end{equation}
where $A$ controls the flow stretching along the vortex axis and the advection of vorticity towards the vortex, $\Gamma$ is the circulation, and $\nu$ is the kinematic viscosity of the air. In our case, if we put $\Gamma = 0.75$~m$^2$/s, and $A = 218$~s$^{-1}$ we obtain $v_\phi^{max} = 200$~m/s, $v_z = 74$~m/s at $z = 18$~cm, which corresponds to the $z-$coordinate of the upper stretcher inlet. Note that this value for $v_z$ is coincident with the airspeed an the inlet of a stretcher, $V_{int} \approx 74$~m/s, estimated from the mass flow in Section \ref{Setup}. With these values for the flow parameters, the resulting vortex diameter is $d = 0.86$~mm. That is $d \sim 1$~mm, roughly speaking. Of course, the Burgers' vortex is a solution for the incompressible Navier-Stokes equations, while a vortex at $M \sim 0.6$ in air is certainly compressible. But we do not expect changes larger than $20$\% in the density or pressure, because our stretching system is not able to sustain pressure drops larger than $20$\% of the atmospheric pressure. Thus, deviation due to compressibility should be less than such percentage.

Due to the turbulent shaking of the filamentary vortex, most of the peaks displayed in Figs.~\ref{four} and \ref{five} are composed typically of two or tree peaks with partial overlapping in time. This effect can be interpreted as the filament going back and forth in the neighborhood of the probe, due to the surrounding turbulence. Fluctuations in the amplitude of peaks can be interpreted as the result of sweepings at different distances from the vortex core, or variations in the Mach number due to varying contents of angular momentum in the incoming flow.
\begin{figure}[t]
\centering \vspace{0.2cm} \hspace{-0.2 cm}
\includegraphics[width=.48\textwidth]{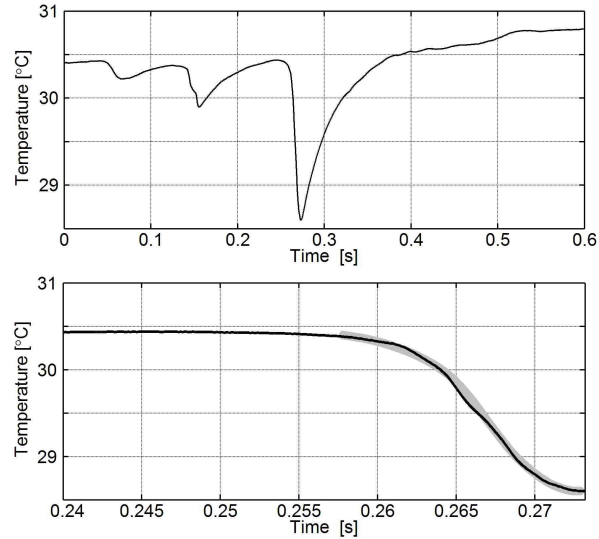}
\caption{ Temperature drop (top) recorded at $d = 1$~cm from the flow axis. Most of these curves are asymmetric (see text).
(Bottom) Expanded view of the falling front. The thick, grey curve in the background is the scaled falling front of
Fig.~\ref{two} (bottom), taken between $x = -2$~cm and $0$, for comparison. The radius of the filament corresponds to the
time interval between $t_1 = 0.264$~s and $t_2 = 0.274$~s. The equivalent core radius in cm cannot be assessed by using
the measurements only (see text).} \label{six}
\end{figure}
All these factors make difficult a clean measurement of speed or temperature profiles. To get a sparser occurrence of events, we located the temperature probe at a greater distance from the flow axis: specifically $r_p = 10$~mm. Fig.\ref{six} (top) displays an isolated event in which the temperature drop was $\Delta T \approx 2 $K. The smaller preceding falls were probably produced by the same vortex, and related to shaking by the turbulence of the surrounding flow while the filament is approaching the probe. Alternatively, periodic drops with increasing amplitude could be due to a helical motion of the filament while it moves towards the probe. Or a combination of both.

As can be seen, the temperature signal is strongly asymmetric: the falling time is much shorter that the recovery time. A possible explanation could be the following: we assume that the interactions between the vortex and the airspeed or temperature probes are different. The former is far more destructive to the filament, because of the hot wire probe structure. In the case of the thermistor, probably its ellipsoidal bead is able to penetrate towards the vortex core without disturbing the flow up to the point of producing a local destruction of its structure. In fact, we found that temperature measurements with the airspeed probe removed are much better, supporting in part this conjecture ---we recall that the distance between probes is only $1.5$~mm---. Once the bead is inside the vortex, the latter tends to remain attached to the thermistor. Thus the bead takes a longer time to leave the vortex core. The tendency of the vortex to remain attached to solid surfaces could also explain the asymmetry of the small temperature falls that precede the large one in Fig.~\ref{six} (top).

Keeping this in mind, we can attempt to get a clue of the temperature distribution inside the filament by looking at the falling edge of the temperature drops. In Fig.~\ref{six} (bottom) we see an expanded view of the falling edge of the largest drop in the curve displayed in Fig.~\ref{six} (top). Clearly it resembles the falling of temperature in the core of the vortex at $M \approx 0.1$. Hence, we have taken the portion between $x = -2$~cm and $0$ of the temperature curve in Fig.~\ref{two} (bottom), and scaled it to make a comparison with the curve in Fig.~\ref{six} (bottom). We see that both curves are similar. Thus, the falling front of the temperature measurement seems to reflect the filament temperature profile, although this assert requires that the vortex translational speed be constant while the probe is traveling towards the core. At this time we cannot assure that, although almost certainly that should not be the case when the probe is leaving the core. If we use the analytical model of reference,\cite{AboVat2008} we conclude that in this specific temperature measurement the Mach number of the filament was  $M = 0.245$.
\begin{figure}[t]
\centering \vspace{0.2cm} \hspace{-0.2 cm}
\includegraphics[width=.48\textwidth]{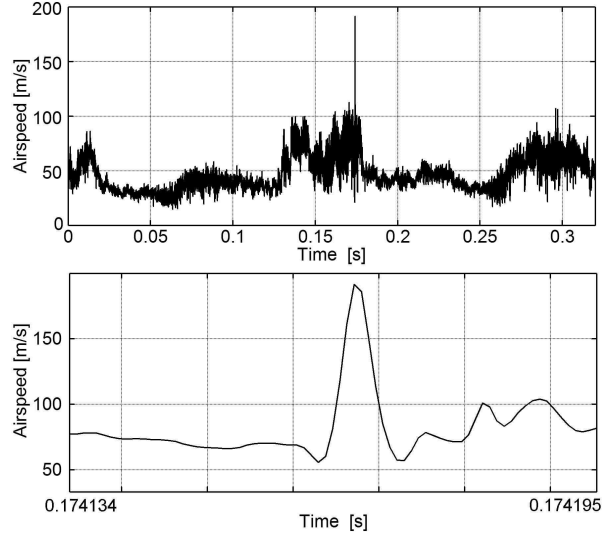}
\caption{(Top) Airspeed measured using a sampling rate of $1.2$~MS/s. The peak at $t \approx 0.17$~s corresponds to an airspeed of $v_t = 191$~m/s.
(Bottom) Expanded view of the airspeed peak. Its width is $\delta t \approx 6~\mu$s . The time scale values are shown only at the ends of the horizontal axis.} \label{seven}
\end{figure}

In the preceding airspeed plots, the sampling frequency for the data acquisition was typically $f_s = 12$~kS/s. To obtain the shape of a peak with enough detail, a sampling rate $f_s = 1.2$~MS/s was necessary. Fig.~\ref{seven} displays an event recorded independently with the probe placed on the flow axis. The value of the peak is $v_t = 191$~m/s, corresponding to $M = 0.56$. Thus, this vortex should produce a temperature drop $\Delta T = 9.85^\circ$C, which is consistent with our independent measurements of temperature. Note that the duration of this event is $t \approx 6$~$\mu$s, which is a very short time as compared with the falling time of a temperature drop, for example, which in the case of Fig.~\ref{six} is $\Delta t \approx 10$~ms, or tree orders of magnitude larger. In fact, if we look at Fig.~\ref{seven} (top) we see that before the peak there is a strong turbulent flow, with maxima reaching $\sim 100$~m/s, and a region of lower speeds near the center. We interpret this as the signal produced by a filament interacting with the hot wire probe and undergoing a local disorganization. When the probe is leaving the filament, there is a flow recovering that produces the large peak that we see in the plot. This happens near the end of the turbulent region, and this pattern is roughly the same in all of the recorded airspeed peaks.

\section{\label{Conc}Conclusions}

To conclude, we remark that the model by Aboelkassem and Vatistas \cite{AboVat2008} gives very good values for the temperature drops in the core of the vortices studied in this work, even though the analytical solution is given for a Prandtl number $Pr = 2/3$, which is about $6$\% smaller than the value $Pr = 0.707$ of the air. The authors state that their result is very insensitive to changes in the Prandtl number: under the conditions of our experiments, deviations not grater than $0.1$\% should be expected as a consequence of this difference in $Pr$. According to the same model, for $M = 0.6$ we could expect pressure drops of about $35$\% of the atmospheric pressure. As we already mentioned, the maximum pressure drop that our stretching system can sustain is about $20$\% of the normal pressure when $Q \rightarrow 0$. Hence, we do not expect pressure drops in the filament core larger than the same fraction of the atmospheric pressure. In fact, with the mass flow that the stretchers must sustain, the inner filament pressure drop should be somewhat less than $20$\% of the atmospheric pressure. Thus, although we did not measured the core pressure directly, by the preceding arguments we believe that the model by Aboelkassem and Vatistas possibly overestimates the pressure drop in the vortex core, despite the very good agreement between the temperature drops given by the theory and those measured in the experiment at corresponding Mach numbers. Finally, we must note that this laboratory free filamentary vortex, with a diameter of about $1$~mm and a length of $34$~cm, have azimuthal velocities as large as $200$~m/s, well above of the airspeeds characterizing the strongest tornadoes observed in nature. On the other hand, in typical turbulent von K\'arm\'an flows with a size similar to that of the region were the filament is formed, the Kolmogorov length is $\eta \sim 300~\mu$m, only about three times smaller than the core diameter. Thus, with the appropriate instruments, this flow offers the possibility of studying the interaction of an extremely strong coherent structure with the small scale turbulent background near the Kolmogorov scale.

\section{\label{Ack}Acknowledgements}

Financial support for this work was provided by FONDECYT under grant No. 1020422, and by DICYT-USACH under project No. 041231LM.

\end{document}